\documentstyle[psbox]{WORKSHOP}

\begin{document}

% TITLE OF THE PAPER
%  If the title is too long for a single line, you can split it 
%  by putting two backslashes. 
%  You might want to put the subtitle. Then it should be inserted 
%  within {\large\sf  }.
%  e.g.:  
%     \title{ Too Long Title \\ for one line \\
%     {\large\sf Subtitle} }
\title{Synergy Between Observations of AGN with GLAST and MAXI\\
{\large\sf  -- Detecting X--ray Precursors of Gamma-ray Flares in Blazars}
}

\author{
Greg Madejski,$^1$ Marek Sikora,$^{1,2}$ and Tsuneyoshi Kamae$^1$
\\[12pt]  % TO BE SPACED WITH ONE LINE
%
% INSTITUTES OF AUTHORS
$^1$  Stanford Linear Accelerator Center, Menlo Park, CA 94025, USA \\
$^2$  Copernicus Center, Warsaw, Poland \\
{\it E-mail(GM): madejski@slac.stanford.edu} 
}

\abst{In five years' time we will witness the launch of two important missions
developed to observe celestial sources in the high energy regime:  
GLAST, sensitive in the high energy $\gamma$--ray band, and MAXI,  
the all-sky X--ray monitor.  Simultaneous monitoring observations by the 
two instruments will be particularly valuable for variable sources, 
allowing cross-correlations of time series between 
the two bands.  We present the anticipated results 
from such observations of active galaxies, and in particular, 
of the jet-dominated sub-class of AGN known as blazars.  
We discuss the constraints on the structure and emission processes 
-- and in particular, on the internal shock models currently invoked 
to explain the particle acceleration processes in blazars -- that 
can be derived with simultaneous $\gamma$--ray and X--ray data.
}

\kword{active galaxies --- blazars --- gamma--rays}

\maketitle
\thispagestyle{empty}

\section{Introduction}

Blazars are active galaxies where the entire observed radiation 
is believed to be dominated by the emission from material moving at 
relativistic speed close to our line of sight.  This is supported, 
among others, by the following observational characteristics:  
compact radio cores, rapid and large 
amplitude variability in all accessible spectral bands, strong 
radio and optical polarization, and spectra described locally as 
power laws.  Blazars are often detected as $\gamma$--ray sources, 
and in some cases, the emission extends up to the TeV range.  
The overall electromagnetic spectra of those sources consist of two 
broad components, one peaking in the infrared -- to -- X--ray range, 
and the other in the $\gamma$--ray range.  The most commonly invoked 
scenarios have the emitting matter moving with relativistic speed 
in a form of multiple clouds or shells, or even as a quasi-continuous 
jet with a Lorentz factor $\Gamma_{\rm jet}$ $\sim 5$ -- 10.  The most viable 
emission mechanisms are the synchrotron process for the low-energy 
component, and Compton upscattering of lower energy photons for 
the high energy component.  Presumably the same population of 
radiating electrons is responsible for both mechanisms;  
the required electron Lorentz factors must reach the range up to 
$10^3 - 10^6$, depending on a particular source.  
Very short radiative lifetimes of such 
particles imply that the acceleration occurs {\sl in situ} -- most 
likely a light-day or more away from the central source.  At least in the 
blazars where strong emission line flux is detected, the most likely source 
of the ``seed'' photons for Compton upscattering is external radiation 
from the broad line region.  
One of the crucial questions regarding astrophysical jets is thus that 
of the conversion of the kinetic energy of the jet to the random 
energy of radiating particles.  

\section {Internal Shock Scenario}

Models for production of nonthermal flares in blazars as well as 
Gamma--ray Burst sources (GRBs) often involve collisions between 
shells containing relatively cold matter, propagating down the 
jet with different velocities.  Such shells approximate 
inhomogeneities which can result from modulation of the relativistic 
outflow by a central engine.  Adopting this model, we demonstrate 
that nonthermal flares, produced by relativistic particles accelerated 
in shocks excited by colliding shells, must be preceded by soft 
X--ray flares, produced by Comptonization of external radiation 
by the material in the cold shells {\sl before the collision}.  
This external radiation is in fact the same source which is 
providing the ``seed'' photons for production of $\gamma$--rays.  
To demonstrate the effect in a simple way, we consider an idealized 
case, where two colliding shells are `symmetrical', i.e. identical 
in their rest frames (equal densities, total masses and negligible
pressures). They are assumed to propagate down the conical jet with
bulk Lorentz factors, $\Gamma_2 > \Gamma_1 \gg 1$, each carrying $N_p$ 
protons.

Provided that the ratio of the electron density $n_e$ to proton density 
$n_p$ obeys $n_e/n_p \ll m_p/m_e$, the protons dominate inertia of the 
shells, and the total energy of two shells before collision is
$$ E = E_1 + E_2 =  N_p (\Gamma_1 + \Gamma_2) m_p c^2 \, .  \eqno(1) $$
From energy and momentum conservation one can find that fraction
of energy dissipated during the collision is
$$ \eta_{diss} \equiv {E_{diss} \over E} =
1 - {2 \Gamma_{1+2} \over \Gamma_1 + \Gamma_2}
\simeq  1- {2 \sqrt{\Gamma_2/\Gamma_1} \over \Gamma_2/\Gamma_1 +1}
\, , \eqno(2) $$
where
$$\Gamma_{1+2} \simeq \sqrt {\Gamma_1 \Gamma_2}   \eqno(3) $$
is the bulk Lorentz factor of the shocked plasma enclosed between the 
forward and reverse shock fronts. Denoting by $\eta_{el}$ the fraction
of dissipated energy  consumed to accelerate relativistic
electrons/positrons, and by $\eta_{rad}$ the average radiative efficiency
of electrons, one can find that the proton content, $N_p$, required
to produce a nonthermal flare with the apparent total luminosity
$L_{fl}$ and lasting $t_{fl}$ is
$$ N_p = {1 \over \eta_{diss} \eta_{el} \eta_{rad}} \,
{L_{fl} t_{fl} \over m_p c^2  {\cal D}^3_{\rm 1+2}} \,
{1 \over (\Gamma_2/\Gamma_1)^{1/2} + (\Gamma_2/\Gamma_1)^{-1/2}}
\, , \eqno(4) $$
where we used Eqs. (1)-(3) and relations:
$$ L_{fl}' = {E_{rad}' \over t_{coll}'} =
{\eta_{diss} \eta_{el} \eta_{rad} E' \over t_{coll}'} \, , \eqno(5) $$
$$ E  = E' \Gamma_{1+2} \, , \eqno(6) $$
$$ t_{fl} = {t_{coll}' \over {\cal D_{\rm 1+2}}} \, , \eqno(7) $$
$$L_{fl} = {\cal D}^4_{\rm 1+2} \, L_{fl}' \, , \eqno(8) $$
and the usual definition of $\cal D$, 
$$ {\cal D}_{\rm 1+2} = {1 \over \Gamma_{\rm 1+2} (1-\beta_{\rm 1+2} \cos \theta_{obs})} \, , \eqno(9) $$
where $t_{coll}'$ is the time scale of the collision, and all primed
quantities are measured in the comoving frame of the shocked plasma.

\begin{figure*}[t]
\centering
%\psbox[xsize=0.4#1,ysize=0.2#1,rotate=r]
\psbox[xsize=15.3cm]
{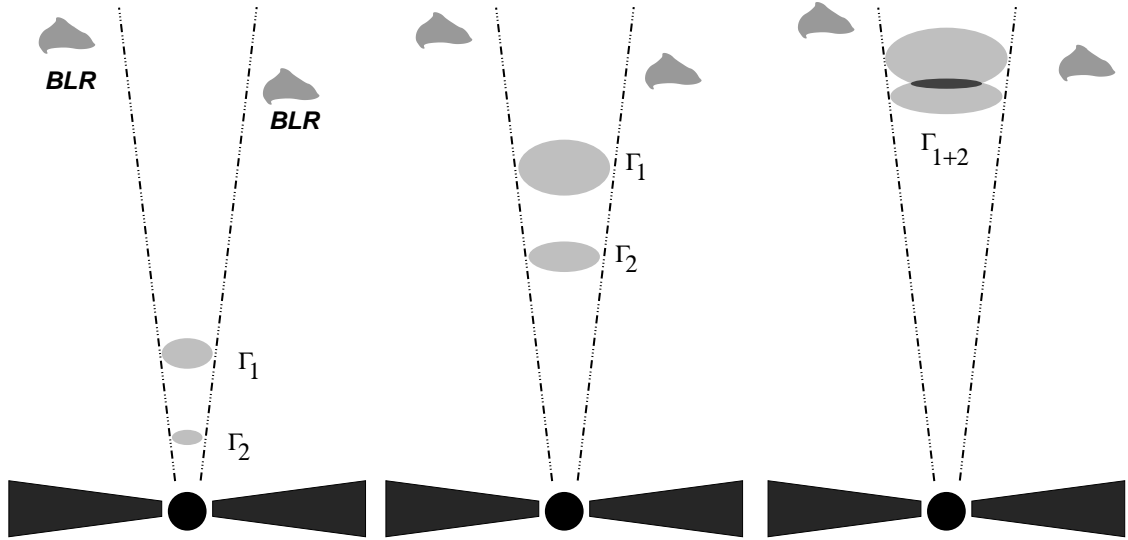}
\caption{Illustration of the scenario for the 
internal shock mechanism discussed in the text, with time progressing 
to the right.  Two cold 
clouds or shells of plasma, moving with Lorentz factors $\Gamma_{1}$ and 
$\Gamma_{2}$ (where $\Gamma_{2} > \Gamma_{1}$) collide, and the 
shocks resulting from the collision provide the site for particle 
acceleration.  BLR denotes the Broad Line Region, which is assumed 
to be the source of ``seed'' photons for Comptonization, resulting 
in X--ray precursors in the left and middle panels, 
and $\gamma$--ray flare in the right panel.  }
\end{figure*}

Number of protons $N_p$ gives us the constraint on the 
minimum number of electrons
contained by each shell. These electrons, which are assumed to be cold before
collision, are predicted to Comptonize external radiation and produce
two spectral features, peaked around frequencies
$$ \nu_i \simeq  {4\over 3} {\cal D}_{\rm i} \Gamma_i \nu_{diff} \, , \eqno(10) $$
where $\nu_{diff}$ is the averaged frequency of the diffuse external
radiation field.  The luminosities of these features are 
$$ L_{SX,i} \simeq N_e \dot E_{el,i} {\cal D}_{\rm i}^4 \, , \eqno(11) $$
$$ \dot E_{el,i} \simeq {4\over 3}\Gamma_i^2 c \sigma_T u_{diff} \, ,
\eqno(12)$$
$$ {\cal D}_{\rm i} = {1 \over \Gamma_i (1-\beta_i \cos \theta_{obs})} \, , \eqno(13)
$$
where $u_{diff}$ is the energy density of the 
diffuse radiation field, and $i = 1, 2$.
The precursors should precede the nonthermal flares by
$$ \delta t_i \sim  {r_{fl,0} \over c} (1-\beta_i \cos \theta_{obs}) \,
\eqno(14) $$
where $r_{fl,0}$ is the distance from the location where the jet was launched 
to that where the cold shells start to collide.  

Adopting the following (fiducial) observables:
$L_{fl} = 10^{48}$ erg s$^{-1}$; $t_{fl} = 1$ day;
$u_{BEL} = 0.03$ erg cm$^{-3}$ (energy density of the 
broad emission line flux);
$\nu_{diff} = 10$ eV, and assuming
$\Gamma_{1+2}=10$, $\Gamma_2/\Gamma_1 =3$ (corresponding to 
$\eta_{diss} \simeq 0.134$),
$\eta_{rad}=0.5$, $\eta_{el}=1/3$
(equipartition between protons, electrons and magnetic fields), and
$\theta_{obs} =1/\Gamma_{1+2}$,
we obtain that the brighter precursor (that produced by the faster shell)
should peak at $\sim 2$ keV and should have luminosity
$$ L_{SX,2} \sim 1.5 \times 10^{45} \,{\rm erg\, \, s}^{-1} {n_e \over n_p}
{u_{diff} \over u_{BEL}} \, , \eqno(14) $$
while a precursor produced by the slower shell should peak around
$\sim 0.7$ keV and should be $(\Gamma_2 / \Gamma_1)^2
({\cal D}_2/ {\cal D}_1)^4 \simeq 9 $ times fainter.  
These preliminary results suggest that during luminous outbursts, 
a brighter precursor should be detected by a moderately sensitive 
detector even if there are no pairs and 
jet formation and collimation is so distant that contribution of
the accretion disc radiation to $u_{diff}$ is negligible.
Assuming that $r_{fl,0} \sim c t_{fl} \Gamma_{1+2}^2 $, the occurrence 
of such soft X--ray precursors should precede 
the bright $\gamma$--ray flares by $\sim 0.7 t_{fl}$ and $\sim 2 t_{fl}$ 
respectively for the brighter and the fainter X--ray precursor.  

\section {Consequences for GLAST and MAXI}

Any studies of correlation of variability in X--rays and $\gamma$--rays 
require simultaneous observations, and it is fortuitous that GLAST 
and MAXI will be operational at the same time.  GLAST is the next generation 
$\gamma$--ray observatory, shown schematically in Fig. 2.  It features 
an effective area given in Fig. 3, which is about 8 times better than 
that of the highly successful EGRET detector 
flown aboard the Compton Gamma--ray Observatory;  
EGRET is of course the detector originally responsible for the 
discovery of $\gamma$--ray flares from many blazars.  
The design of GLAST allows monitoring simultaneously of more 
than 2 steradians of the sky, so such a large number of such 
flares will be detected and monitored from multiple 
blazars at the same time.  

In contrast, most current X--ray detectors 
feature a relatively narrow field of view, typically less 
than $1 \times 1$ degree.  It is of course possible to train 
an X--ray observatory on a blazar discovered to be flaring 
in $\gamma$--rays, but the crucial observation is the history of
X--ray flux {\sl prior} to such a flare:  with this, the monitoring of the 
entire sky in the X--ray band is required.  
Without an X--ray monitoring instrument which would detect such 
a flare {\sl before} it is measured by GLAST, one would have to rely 
on a rare coincidence that an X--ray telescope 
would be observing a blazar which is about to flare 
in the $\gamma$--ray band.  
These preliminary results suggest that during luminous outbursts, 
at least relatively bright precursors should be detected 
by an all-sky sensitive instrument such as MAXI.  Finally, 
its energy resolution also plays an important role in 
making detailed estimates of the physical conditions in the ``early'' 
phases of the formation of jets in blazars.  

\begin{figure*}[t]
\centering
%\psbox[xsize=0.4#1,ysize=0.2#1,rotate=r]
\psbox[xsize=10cm]
{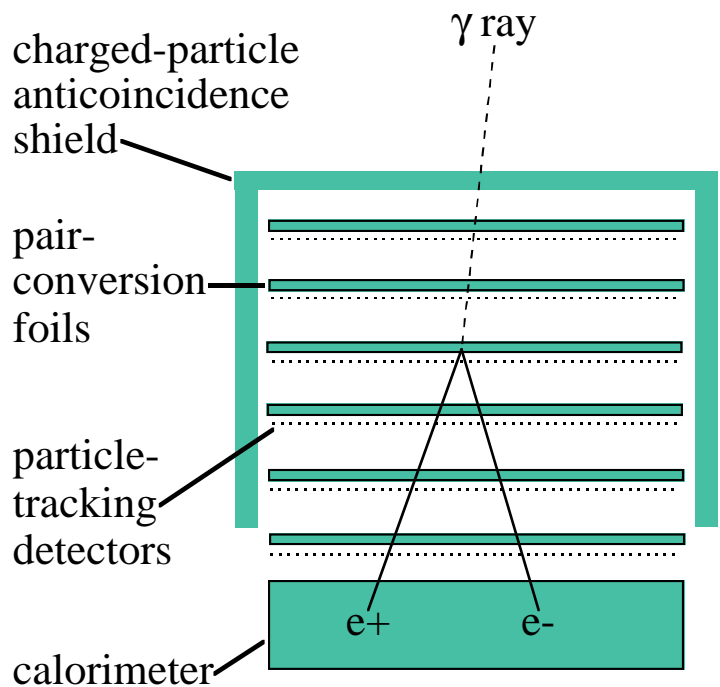}
\caption{Schematic illustration and the principle of operation of the 
GLAST Large Area Telescope (LAT).  Entering $\gamma$--ray converts into an 
e$^+$ / e$^-$ pair in the conversion foil.  The resultant e$^+$ and e$^-$
charges deposit energy and are detected in the particle-tracking detectors, 
which are made of position-sensitive silicon strips.  
Recording of the locations where the charges interacted with 
the silicon strips allows a reconstruction of their ``tracks.''  
This in turn permits the determination of the 
direction of arrival of the incident $\gamma$--ray.  The calorimeter, 
located beneath the tracker, is a scintillator, capable of measuring 
the total energy deposited by the particles produced by the 
incident $\gamma$--ray.  The LAT detector is modular:  it consists of 
a $4 \times 4$ array of identical towers.  
The entire detector system is enclosed by 
an anti-coincidence shield, designed to reject the charged 
particle background.  
}
\end{figure*}

\begin{figure*}[t]
\centering
%\psbox[xsize=0.4#1,ysize=0.2#1,rotate=r]
\psbox[xsize=12cm]
{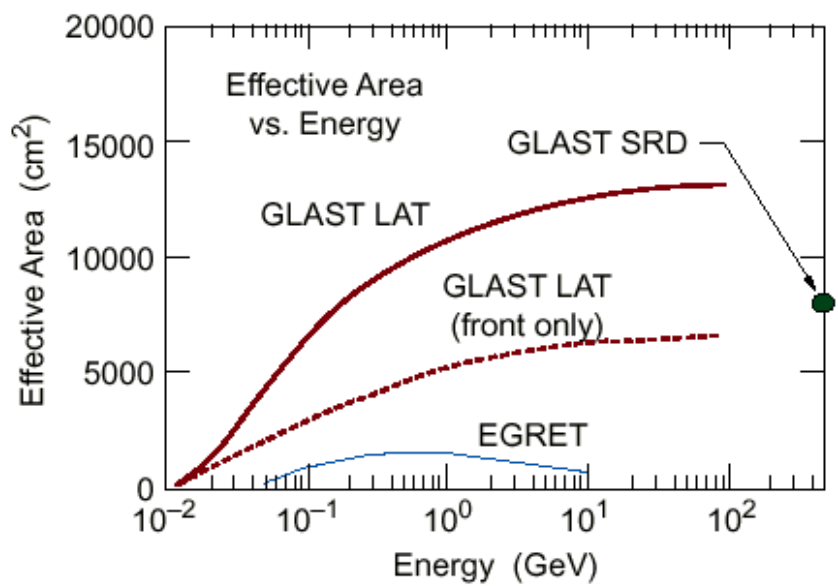}
\caption{Effective area of the GLAST LAT instrument as a function of
energy.  The instrument has some sensitivity at 20 MeV;  its 
effective area is greatest between $\sim 0.1$ and 100 GeV.  One of 
the design criteria of the instrument was to extend the energy bandpass 
to at least 100 GeV, to allow some overlap with ground-based TeV 
Cerenkov air shower array detectors.  Current plans for GLAST have the 
sensitivity extending to the point marked as ``GLAST SRD,'' where 
even the current TeV detectors have a reasonably good sensitivity. }
\end{figure*}

We would like to emphasize that our predictions regarding 
soft X--ray precursors 
are independent on the specific model of nonthermal radiation mechanism.
Eventual lack of precursors can jeopardize
the internal shock models and favor the external, reconfinment
shocks and/or reconnection of magnetic fields as the source of energy
for relativistic electrons.  In case of positive
detection, the detailed analyses can be used to trace the
structure of innermost parts of relativistic jets and to put constraints
on a pair content. So far, only observations of 3C 279 provided 
sufficiently good data to allow to search for such precursor 
(Wehrle et al. 1998; Lawson, McHardy \& Marscher 1999).  
Detailed analyses of the X--ray spectral index evolution before 
and during the flare show possible presence of precursor:  
the modest flux detected from this flare 
excludes high pair content in this object.
We also note that studies of Comptonization of external 
radiation by cold electrons in a jet were performed previously 
(see, e.g., Begelman \& Sikora 1987 and
Sikora, Madejski, Moderski, \& Poutanen (1997), but those considered 
only stationary jets.  Consideration of such stationary 
case makes it difficult to distinguish
the contribution of precursors originating from 
the bulk Compton radiation described above from other sources of 
soft X--rays.  

\section{Acknowledgements}

We acknowledge the support from NASA Chandra observing grant GO0-1038A 
and the Polish KBN grant 2P03D 00415, and the help of Dr. Berrie Giebels 
with the figures.  MS is grateful for the 
hospitality of SLAC, where the research reported here was conducted.  

\section{References}

\re 
Begelman, M., \& Sikora, M. 1987, ApJ, 322 650.  

\re 
Lawson, A., McHardy, I., \& Marscher, A. 1999, MNRAS, 306, 247.

\re 
Sikora, M. Begelman, M., \& Rees, M. 1994, ApJ, 421, 153.  

\re
Sikora, M., Madejski, G. M., Moderski, R., \& Poutanen, J. 
1997, ApJ, 484, 108.  

\re 
Wehrle, A., et al. 1998, ApJ, 497, 178.

\label{last}

\end{document}